\documentclass[reprint,showpacs,preprintnumbers,amsmath,amssymb,a4paper,nofootinbib,superscriptaddress,floatfix,aps,prb,longbibliography,showkeys]{revtex4-1}

\usepackage[mathlines]{lineno}
\usepackage{blindtext, color}

\usepackage{graphicx}

\usepackage{dcolumn}
\usepackage{bm}
\usepackage{amsmath}
\usepackage[color=green]{todonotes}
\usepackage{placeins} 

\usepackage{color}
\usepackage{colortbl}

\usepackage{txfonts}

\usepackage{grffile}

\makeatother

\begin{document}

\preprint{$\today$}

\title{Heisenberg pseudo-exchange and emergent anisotropies in field-driven pinwheel artificial spin ice}

\author{Gary W. Paterson}
    \email{gary.paterson@glasgow.ac.uk}
    \affiliation{SUPA, School of Physics and Astronomy, University of Glasgow, Glasgow, G12 8QQ, UK}
\author{Gavin M. Macauley}
    \affiliation{SUPA, School of Physics and Astronomy, University of Glasgow, Glasgow, G12 8QQ, UK}
\author{Yue Li}
    \altaffiliation[Present address: ]{Materials Science Division, Argonne National Laboratory, Lemont, IL, 60439, USA}
    \affiliation{SUPA, School of Physics and Astronomy, University of Glasgow, Glasgow, G12 8QQ, UK}
\author{Rair Mac\^edo}
    \affiliation{James Watt School of Engineering, Electronics \& Nanoscale Engineering Division, University of Glasgow, Glasgow, G12 8QQ, UK}
\author{Ciaran Ferguson}
    \affiliation{SUPA, School of Physics and Astronomy, University of Glasgow, Glasgow, G12 8QQ, UK}
\author{Sophie A. Morley}
    \altaffiliation[Present address: ]{Lawrence Berkeley National Laboratory, Berkeley, CA 94720, USA}    \affiliation{School of Physics and Astronomy, University of Leeds, Leeds, LS2 9JT, UK}
\author{Mark C. Rosamond}
    \affiliation{School of Electronic and Electrical Engineering, University of Leeds, Leeds, LS2 9JT, UK}
\author{Edmund H. Linfield}
    \affiliation{School of Electronic and Electrical Engineering, University of Leeds, Leeds, LS2 9JT, UK}
\author{Christopher H. Marrows}
    \affiliation{School of Physics and Astronomy, University of Leeds, Leeds, LS2 9JT, UK}
\author{Robert L. Stamps}
    \affiliation{Department of Physics and Astronomy, University of Manitoba, Winnipeg, MB R3T 2N2, Canada}
\author{Stephen McVitie}
    \affiliation{SUPA, School of Physics and Astronomy, University of Glasgow, Glasgow, G12 8QQ, UK}

\date{\today}

\begin{abstract}
Rotating all islands in square artificial spin ice (ASI) uniformly about their centres gives rise to the recently reported pinwheel ASI.
At angles around 45$^\mathrm{o}$, the antiferromagnetic ordering changes to ferromagnetic and the magnetic configurations of the system exhibit near-degeneracy, making it particularly sensitive to small perturbations.
We investigate through micromagnetic modelling the influence of dipolar fields produced by physically extended islands in field-driven magnetisation processes in pinwheel arrays, and compare the results to hysteresis experiments performed \emph{in-situ} using Lorentz transmission electron microscopy.
We find that magnetisation end-states induce a Heisenberg pseudo-exchange interaction that governs both the inter-island coupling and the resultant array reversal process.
Symmetry reduction gives rise to anisotropies and array-corner mediated avalanche reversals through a cascade of nearest-neighbour (NN) islands.
The symmetries of the anisotropy axes are related to those of the geometrical array but are misaligned to the array axes as a result of the correlated interactions between neighbouring islands.
The NN dipolar coupling is reduced by decreasing the island size and, using this property, we track the transition from the strongly coupled regime towards the pure point dipole one and observe modification of the ferromagnetic array reversal process.
Our results shed light on important aspects of the interactions in pinwheel ASI, and demonstrate a mechanism by which their properties may be tuned for use in a range of fundamental research and spintronic applications.
\end{abstract}


\maketitle

\section{Introduction}
The patterning of materials on the nanoscale through modern fabrication techniques has enabled the creation of arrays of interacting magnetic islands, named artificial spin ice~\cite{Wang_nature_2006_square_ASI} (ASI) after the canonical example of water ice in which geometrical frustration exists.~\cite{Pauling_jacs_35_ice}
In these arrays, each island is typically elongated so that a single magnetic domain is formed within the island and constrained to lie along its long-axis.
The so-called Ising~\cite{Ising_1925em} macrospins represented by each island act as analogues of classical magnetic spins, allowing insight into real atomic systems.~\cite{Harris_PRL_1997_pyrochlore}
The lateral dimensions of each island, of the order of 10s to 100s of nanometres, and their geometrical layout can be tailored to alter the local inter-island interactions with profound affects on the properties of the system.~\cite{Mengotti_PRB_2008_kagome, Morrison_njop_2013_frustration_geom, Nisoli_RMP_2019_rev}
The collective behaviour of the systems and their tunability makes promising their potential use in fundamental physics research, and a wide range of spintronic and magnonic applications.~\cite{Heyderman_JPCM_2013_ferroic_systems}

\begin{figure*}[hbt]
  \centering
      \includegraphics[width=17.5cm]{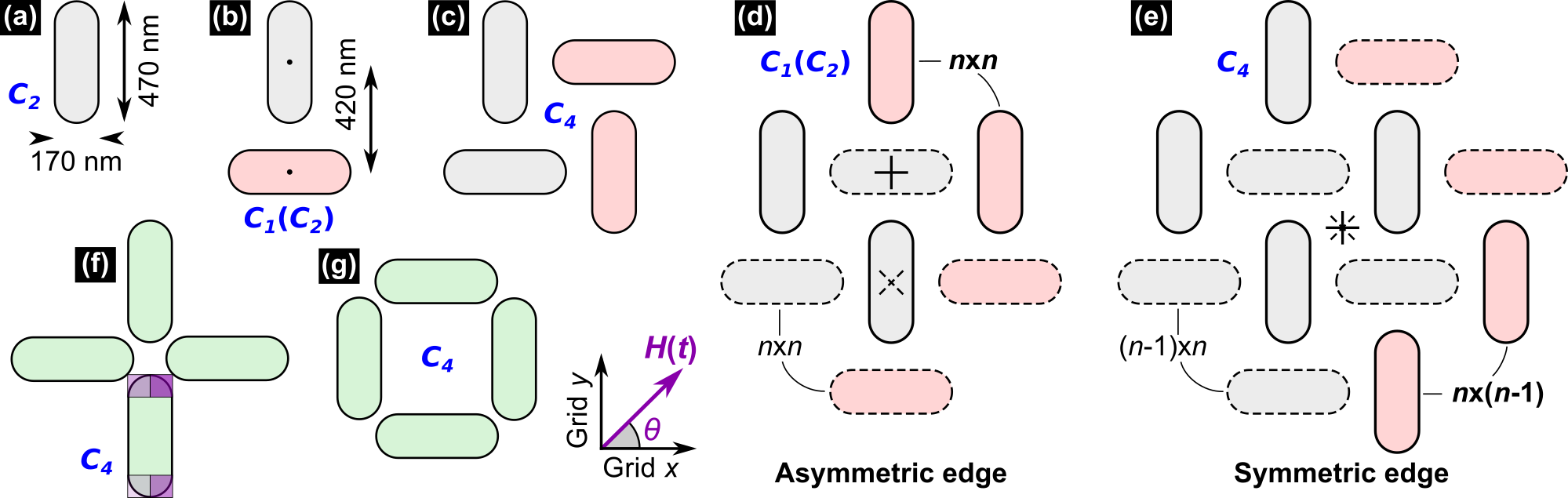}
      \caption{
        ASI geometries considered. (a) A single island, (b) two islands forming a `T-shape', (c) a pinwheel `unit', and larger (d) `asymmetric' and (e) `symmetric' pinwheel arrays. Square ASI (f) `vertex' and (g) `unit'. All `units' and `vertices' are referred to as the latter in the main text. In (a)-(e) the red islands are those added to the previous structure to generate the current one. The solid and dashed crosses in (d) and (e) show the centres of the subarrays with islands outlined in the corresponding lines. These sets of islands form subarrays of $n{\times}n$, ($n-1$)${\times}n$, or $n{\times}$($n-1$) islands, as annotated. The blue ($C_x$) fonts show the 2-D rotational symmetry of the geometry. The same parameter is displayed in brackets when reciprocity of the field process increases the symmetry. The inset to (g) shows the applied field orientation definition and the simulation grid axes. The overlays in the bottom island in (f) show the regions used to track the end-states of each island. The annotations in (d) and (e) refer to the subarray repeat shapes for the two edge symmetry types.
      }
      \label{fig:asi_geom}
\end{figure*}

The square ASI geometry has an antiferromagnetic (AFM) ground state (GS) ordering, where each set of four islands is formed by two orthogonal pairs of collinear Ising spins lying in a plane, with one pair aligned head-to-head and the other pair tail-to-tail.~\cite{Wang_nature_2006_square_ASI}
In square ASI, the GS is well defined,~\cite{morgan_NatPhys_2010_sq_AFM_GS} with the higher energy states separated significantly in energy,~\cite{Wang_nature_2006_square_ASI, Macedo_PRB_2018} and field-driven reversal occurs through sequential chain flipping~\cite{Morgan_njop_2011_sq_reversal, Phatak_PRB_2011_sq_monopole_edge} also referred to as Dirac strings.
The recently reported `pinwheel' geometry~\cite{Gliga_nmat_2017, Macedo_PRB_2018, Li_acs_nano_2019, Macauley_2019} is created by rotating each island in a square lattice about its centre by 45$^\mathrm{o}$ (c.f. Figs.~\ref{fig:asi_geom}(f) and \ref{fig:asi_geom}(c)).
Within the contexts of point-dipole and micromagnetic models, rotation of the islands has been predicted to modify the inter-island coupling in very similar ways.
As the island rotation angle is increased from 0 to 45$^\mathrm{o}$, the dominant nearest-neighbour (NN) coupling in square ASI decreases in favour of an increased coupling to more distant islands.~\cite{Macedo_PRB_2018}
In a small range of rotation angles around 45$^\mathrm{o}$, the energy level spacings are significantly reduced, creating a near-degenerate system with two-dimensional (2-D) superferromagnetic~\cite{Bedanta_PRL_2017_superFM} GS ordering.~\cite{Macedo_PRB_2018}

The transition between square and pinwheel ice has been mapped as a function of rotation angle, yielding insight into defect formation and the demonstration of a true ice manifold in a 2-D system.~\cite{Macauley_2019}
Additionally, thermally-driven magnetisation reversal processes have been observed and attributed to emergent chiral dynamics.~\cite{Gliga_nmat_2017}
Ferromagnetic (FM) ordering of pinwheel ASI has been observed in field-driven experiments, with the system exhibiting unusual charged domain walls, configurable one-dimensional (1-D) and 2-D reversals, and an unexpected misalignment of the anisotropy axis with respect to the array axes.~\cite{Li_acs_nano_2019}
These field-driven properties, along with the magnetisation processes itself, could not be explained by Monte Carlo simulations using a point-dipole model to represent each island macrospin.

It is commonly assumed that the macrospin model holds true in ASI samples as each island is sufficiently small to support a single domain ground state.
While single domains are often observed in well-designed experiments, the extended size of the islands and the deviation from the macrospin model through curvature of the magnetisation at the edges of the islands in `end-states' to minimise energy are often overlooked.
End-states are generally difficult to observe experimentally, but have been seen in square ice through holography.~\cite{Phatak_PRB_2011_sq_monopole_edge}
In simulations, when comparisons have been made, it is generally for zero-field conditions and little difference is found.~\cite{Macedo_PRB_2018, Louis_NatMat_2018_potts}
However, end-states have been shown to be important in defining a chirality for monopoles in kagome ASI, giving rise to anisotropic reversal;~\cite{Rougemaille_NJP_2013_chiral_end_states} in determining microwave frequency dynamics of square~\cite{Gliga_PRB_2015_sq_vertex_endstate} and Kagome~\cite{PANAGIOTOPOULOS_jmmm_2017_kagome_RF} ASI; in the systematic creation of vortex flux closure states in coupled islands;~\cite{Porro_JAP_2012_vortex_islands} and in altering the reversal fields in coupled 1-D island chains.~\cite{Nguyen_PRB_2017_1d_phase_diag_end_states}
In pinwheel ASI, the reduction in the interaction strengths means that this system aught to be particularly sensitive to modifications to the dipolar coupling due to the presence of end-states.

In this paper, we investigate the influence of dipolar fields arising from extended islands in pinwheel ASI through micromagnetic modelling and compare our results to experiment.~\cite{Li_acs_nano_2019, Li_acs_nano_2019_enlighten}
We find that dipolar interaction between NN islands induces a Heisenberg pseudo-exchange effect which creates a strongly coupled regime associated with 2-D superferromagnetism.
Within the regime, spatial inhomogeneity in the island switching fields arises due to the reduced symmetry of the arrays at their edges, resulting in array-corner mediated avalanche reversals.
Related to this are emergent cubic and uniaxial anisotropy contributions to the energy landscape.
The nature of the array reversal and anisotropies can be modified by varying the island size, allowing tuning of the array properties for potential use in a wide range of fundamental research and applications, such as Hall circuits,~\cite{Chern_pra_17_hall_circuits} logic,~\cite{Imre_science_2006_ASI_computation, Arava_2018_sq_logic, Caravelli_arxiv_2018_bool_gates} and neuromorphic computation.~\cite{Jensen18_asi_computation, Arava_PRA_2019_asi_relax}

\section{Simulated geometries}
Micromagnetic simulations were performed across the geometries shown in Fig.~\ref{fig:asi_geom} and with larger arrays of the same type.
The islands were 470~nm $\times$ 170~nm in lateral size and 10~nm thick, with a spacing between island centres of 420~nm, as shown in Figs.~\ref{fig:asi_geom}(a) and \ref{fig:asi_geom}(b), matching the nominal values of our earlier work.~\cite{Li_acs_nano_2019}
The single island in Fig.~\ref{fig:asi_geom}(a) is used as a reference and has $C_2$ rotational symmetry, as indicated by the blue text in the figure.
In each of Figs.~\ref{fig:asi_geom}(a) to \ref{fig:asi_geom}(e), we add an additional layer of islands, shown in red, to the previous panel, enlarging the array while maintaining as equal the number of horizontal and vertical islands.
The two island configuration in Fig.~\ref{fig:asi_geom}(b) forms an inverted `T-shape' and is the simplest geometry in which to investigate extended island effects.
This T-shaped geometry has $C_1$ rotational symmetry.

The structures in Fig.~\ref{fig:asi_geom}(b)-\ref{fig:asi_geom}(e) are of different sized pinwheel arrays.
These are formed by two interleaved subarrays of collinear islands with a 90$^\mathrm{o}$ offset between the islands of the two subarrays.
We restrict our geometry to arrays formed from either two $n{\times}n$ subarrays [Fig.~\ref{fig:asi_geom}(d)] or by a $n{\times}$($n$-1) subarray interleaved with a ($n$-1)${\times}n$ subarray [Figs.~\ref{fig:asi_geom}(c) and \ref{fig:asi_geom}(e)].
Each subarray has $C_2$ symmetry, but the rotational symmetry of the overall arrays alternates between $C_4$ and $C_1$ as a result of the subarray centres being aligned or offset from one another (see crosses in Figs.~\ref{fig:asi_geom}(d) and \ref{fig:asi_geom}(e)), creating arrays with `symmetric' or `asymmetric' edges.

The four-island pinwheel array shown in Fig.~\ref{fig:asi_geom}(c) has a symmetric edge and is the smallest lattice forming a 2-D array.
Earlier work~\cite{Macedo_PRB_2018} termed this structure a `unit', to differentiate it from the `vertex' commonly used in square ASI, where a set of four islands meet head on, as shown in Fig.~\ref{fig:asi_geom}(f).
In this paper, we will simply refer to all sets of four islands as vertices.

All simulations were performed using the GPU-accelerated MuMax3 package.~\cite{mumax3_2014}
Since we are primarily interested in field-driven processes, we mainly use the magnetisation switching field values as proxies to the \emph{net} interactions.
The system was initialised with the magnetisation saturated parallel to an external field of -100~mT, applied at an angle $\theta$ with respect to the $x$-axis, as shown in the inset to Fig.~\ref{fig:asi_geom}(g).
The switching field, $H_s$, was determined by the point at which the magnetisation along the easy-axis of each island reversed.

To map the anisotropies present, the simulations were repeated at multiple applied field angles, generating switching field `astroids'.
These astroids mark the boundary between regions of single and double island energy minima in the range of coherent magnetisation rotation, and have been used to characterise other single domain structures.~\cite{Thiaville_PRB_2000_astroids}

The end-states of each island were also tracked by recording the mean magnetisation in the four `corners' of each island, as indicated by the overlays to the bottom island in Fig.~\ref{fig:asi_geom}(f).
A measure of the end-state strength was made by taking the mean of the product of a matching template with the component of the reduced magnetisation perpendicular to the island long-axis in the four regions.
In this figure of merit, end-states with more (less) curvature away from the long-axis of the island have greater (smaller) strength.

Although the magnetisation process itself is non-reciprocal, since the magnitude of the switching field values should be equal on each side of the hysteresis loop, the symmetry of the astroids is increased for the asymmetric arrays. 
For example, in the T-shaped array, the symmetry is increased from $C_1$ to $C_2$.
The lowest expected symmetry of the switching field across all structures is $C_2$, and we therefore restrict the simulation range to [0$^\mathrm{o}$, 180$^\mathrm{o}$].

To track details of the reversal, time resolved simulations were performed which allow the magnetisation to evolve according to the full equation of motion.
These simulations show that spin waves occur within the islands after reversal, and perturb the stray field.
However, we see no evidence of dynamics influencing the reversal properties.~\cite{sup}

Unless otherwise stated, in the simulations, the angle and field step resolutions at angles around the anisotropy axes and fields around the switching field were 0.25$^\mathrm{o}$ and 25~$\muup$T, respectively; the in- and out-of-plane cell sizes were 5~nm and 10~nm, respectively; the exchange constant was 13~pJ/m; the saturation magnetisation was 800~kA/m; and the damping was set to 0.02.
Further details of the simulations can be found in Supplemental Material \S1.~\cite{sup}
All data was processed with Python and the FPD library.~\cite{fpd_library}

\section{Zero and one dimensional arrays}
\begin{figure}[hbt]
  \centering
      \includegraphics[width=8.5cm]{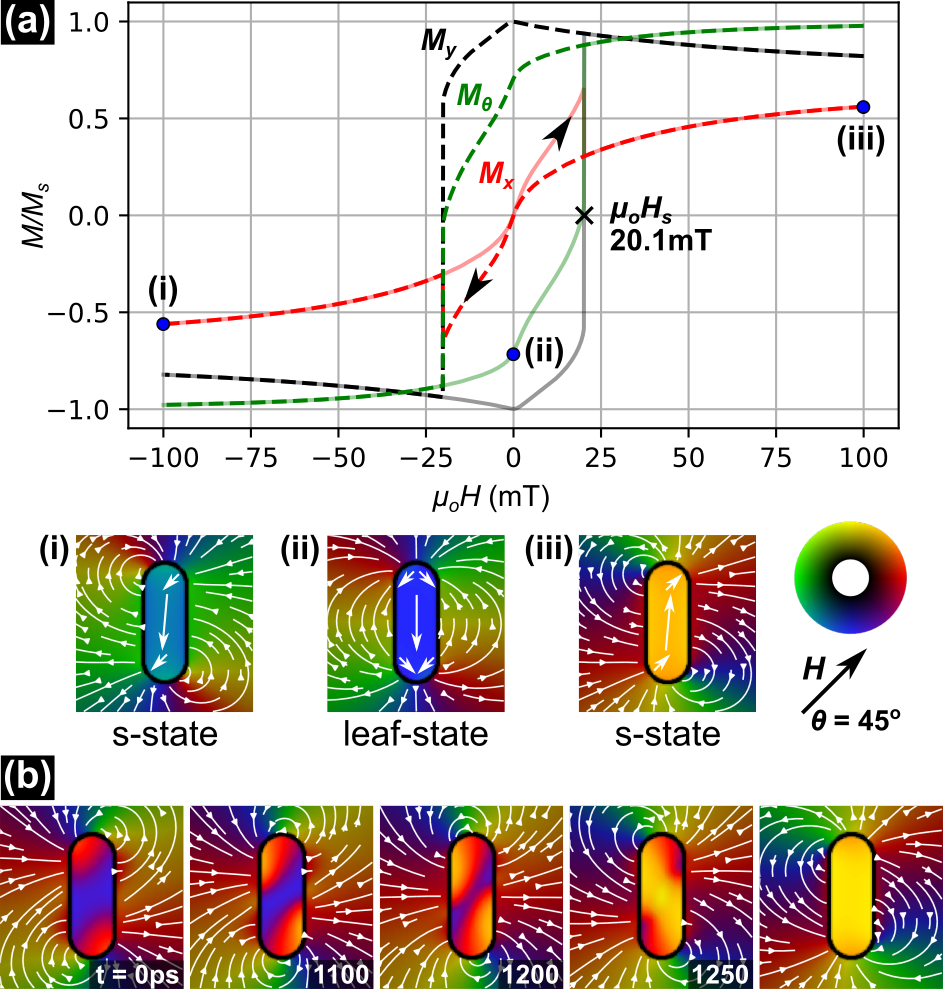}
      \caption{
        Single island magnetic properties with $H$ applied at 45$^\mathrm{o}$. (a) $M$-$H$ loop, showing magnetisation configurations at key points (i) - (iii) and the switching field by the annotation. The solid (dashed) lines show the increasing (decreasing) field segments, as indicated by the arrows. (b) Time resolved end-state mediated magnetisation reversal. The last panel depicts the steady state configuration. The stray field is plotted on a logarithmic scale. The colour wheel in (a) represents the orientations and relative magnitudes of the vectors. The white arrows within the islands are sketches to aid the eye and are not quantitative. 
      }
      \label{fig:0d_mh}
\end{figure}

Fig.~\ref{fig:0d_mh}(a) shows the $M$-$H$ loop of a single vertical island, with the external field applied at an angle of 45$^\mathrm{o}$.
The components parallel ($M_y$, black) and perpendicular ($M_x$, red) to the island long-axis, and that along the applied field direction ($M_{\theta}$, green) are shown.
As noted previously, we take the switching field, $H_s$, as that at which the magnetisation along the long-axis of the island reverses and, as shown in the figure, this is marked by a sharp transition.

\begin{figure}[hbt]
  \centering
      \includegraphics[width=8.5cm]{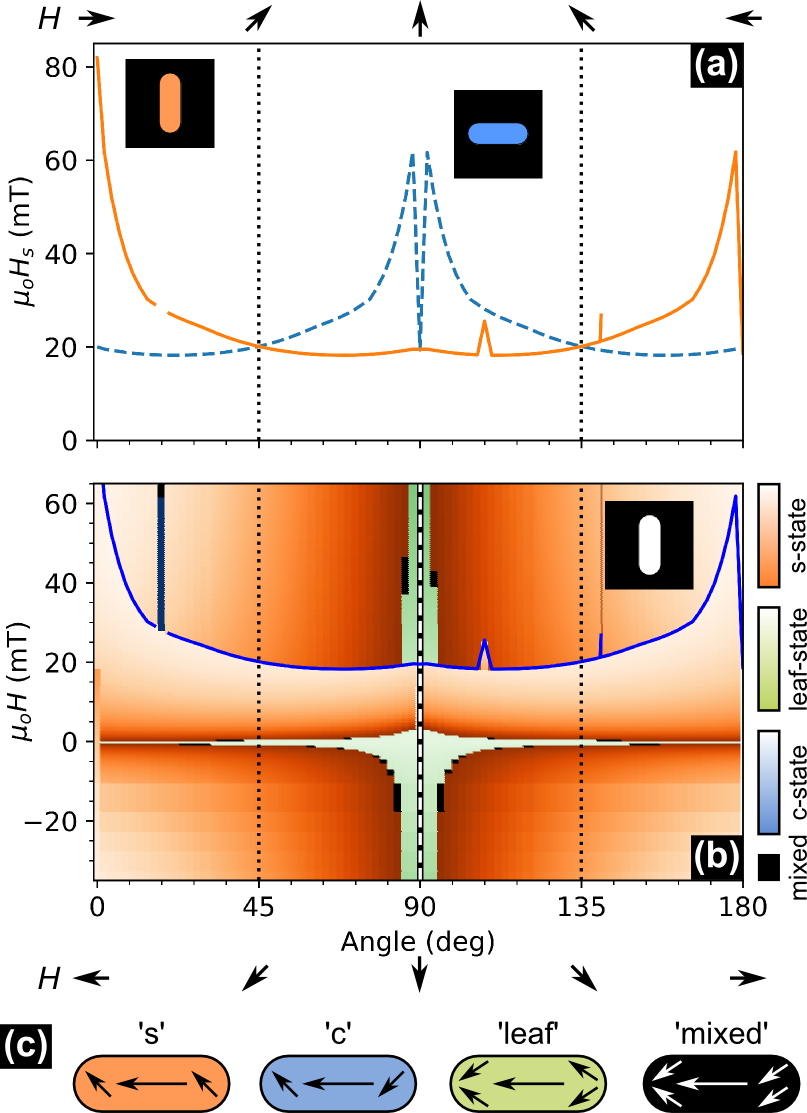}
      \caption{
        Magnetic property angular dependence of a single island. (a) Applied field angle dependence of $H_s$ for a vertical and horizontal island. (b) Vertical island end-state phase diagram with $H_s(\theta)$ superimposed (line). For all pure states, increasing lightness marks a stronger component perpendicular to the long-axis of the island, i.e. a stronger end-state. The color bars are normalised for each end-state between zero and the maximum value. (c) Example end-state schematics. The arrows in (a) and (b) represent the field directions at positive (top) and negative (bottom) fields. 
      }
      \label{fig:0d_angles}
\end{figure}

Panels (i) to (iii) of Fig.~\ref{fig:0d_mh}(a) show example magnetisation configurations during the increasing field sweep at the locations similarly marked in the $M$-$H$ loop.
The island is single domain with s-type end-states at field (i) and (iii), and a weak leaf-state at remanence (ii).
In leaf-states (sometimes called onion states), the magnetisation follows to some extent the shape of the island to reduce the contribution to the demagnetizing energy.
Example schematics of different end-states are shown in Fig.~\ref{fig:0d_angles}(c).
In Fig.~\ref{fig:0d_mh}(a) and throughout this work, the magnitude and orientation of the vector is defined by the inset colour wheel.
The stray field is plotted with the same colour map, but on a logarithmic scale.
The black borders on the islands here and elsewhere are drawn for clarity.

Details of the reversal mechanism are shown in the time resolved magnetisation maps of Fig.~\ref{fig:0d_mh}(b), where the applied field is increased from just below to just above $H_s$.
The s-states grow symmetrically in strength and extent, eventually extending down the sides of the island, after which the centre of the island reverses in a coherent rotation.
We emphasise that the end-states shown here are a primary contributory to the $x$-component of the magnetisation in Fig.~\ref{fig:0d_mh}(a) and is markedly different to the macrospin or point-dipole models, where no end-states exist and the $M$-$H$ loop would be a perfect step function.

For the applied field angle of 45$^\mathrm{o}$ shown in Fig.~\ref{fig:0d_mh}, reversal is through an anticlockwise rotation.
Indeed, the island acts as a spin-ratchet during field-driven reversal at any applied field angle not equal to the (shape) anisotropy axis, with the sense of rotation depending on the relative angle. 
For example, if the angle were changed to be within (90$^\mathrm{o}$, 180$^\mathrm{o}$], then reversal of the magnetisation would be through clockwise rotation.
We will return to this property of anisotropy axes later, when considering spin ice arrays.
Between the time of 1250~ps and the steady state configuration shown in the last panel of Fig.~\ref{fig:0d_mh}(b), spin waves reflect off the walls of the island as it reaches equilibrium (see Supplemental Material Video~S1~\cite{sup}). 
Reversal is from one s-state to another and is mediated by the end-states, altering the stray field and, thus, the coupling to any adjacent islands.

Fig.~\ref{fig:0d_angles}(a) shows the dependence of $H_s$ on the applied field angle $\theta$, for an isolated vertical island (solid line) and a horizontal one (dashed line).
The two astroids are the same except for a 90$^\mathrm{o}$ shift,
\footnote{%
  We note that the `noise' in the $H_s(\theta)$ plots is the result of metastable states being encountered in the simulations, as indicated by the coincidence of the same `noise' in the end-state phase diagrams and in the total energy density (shown in Supplemental Material Fig.~S3~\cite{sup}).
  While we could identify and repeat or initialise the simulation under different conditions until the ground state is found in all cases, we instead represent the data as it was obtained from a single set of simulations.
  We do this in order to show that the main features of the results we present are almost unaffected by a small number of metastable states, in agreement with the previously reported related experiment~\cite{Li_acs_nano_2019} in which there was also evidence of a small number of metastable states across the array.
}
and peak with the field applied along the island hard-axes, as expected.
A slight peak is seen in $H_s(\theta)$ when $H$ is applied along the easy-axis due to the existence of leaf end-states, as the sense of rotation of the end-states which mediate reversal is less well established.

The end-state phase diagram for a single vertical island is shown in Fig.~\ref{fig:0d_angles}(b).
For all pure states, increasing lightness marks a stronger component perpendicular to the long-axis of the island, i.e. a stronger end-state.
The majority of end-states are s-type (orange), with some leaf states (green) when the field lies close to the long-axis, and also when the field is close to zero.
Mixed end-states, combinations of c-, s- and leaf-type, occur at the transitions between s-type and leaf end-states and are depicted in black.

\begin{figure}[hbt]
  \centering
      \includegraphics[width=8.5cm]{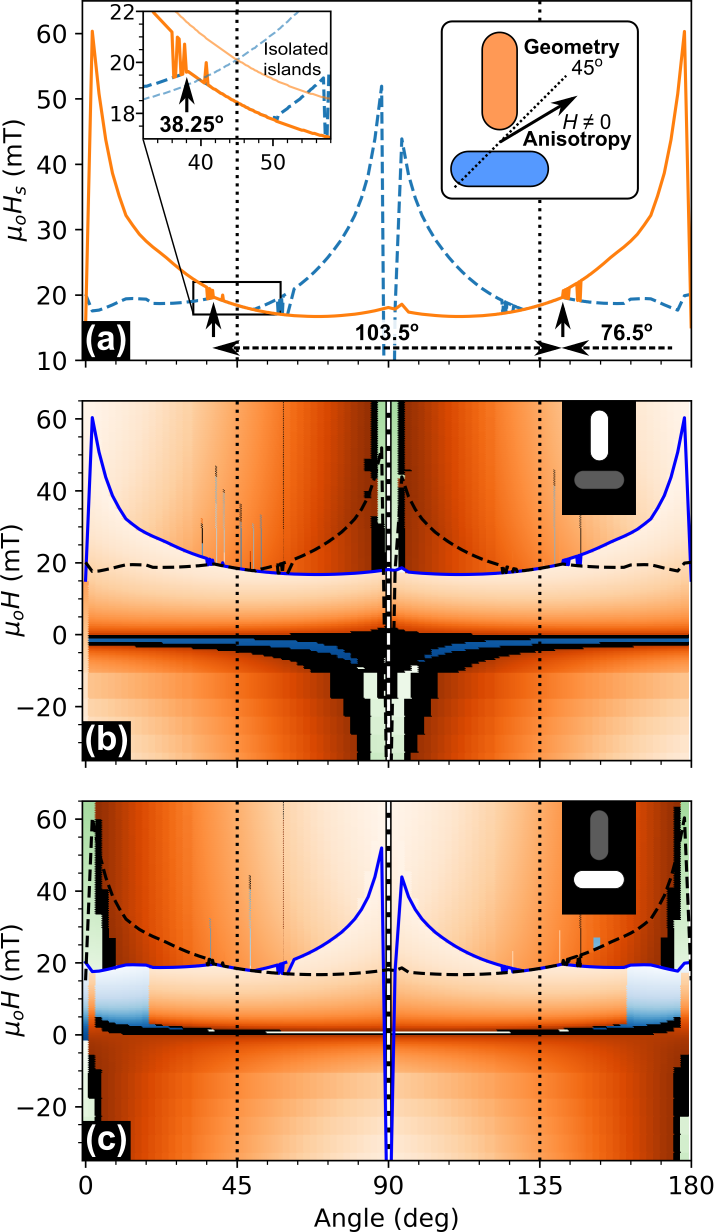}
      \caption{
        Magnetic properties of the T-shaped geometry. (a) $H_s(\theta)$ astroids. The thin lines in the inset to (a) are the equivalent lines for the isolated islands of Fig.~\ref{fig:0d_angles}(a). (b) Vertical and (c) horizontal island end-state phase diagrams. The colour maps in (b) and (c) are the same as those presented in Fig.~\ref{fig:0d_angles}(c). The solid lines in the phase diagrams are the superimposed $H_s(\theta)$ profiles of the highlighted island. The dashed lines are for the other island highlighted in (a).
      }
      \label{fig:1d}
\end{figure}

\begin{figure*}[hbt]
  \centering
      \includegraphics[width=17cm]{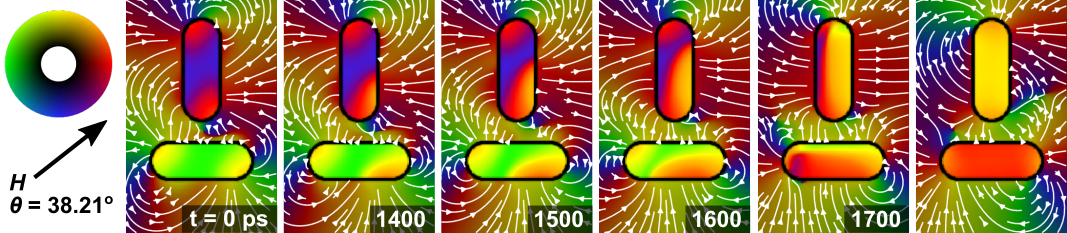}
      \caption{
        Time evolution of the magnetisation and stray field of the two-island T-shaped geometry during reversal at the anisotropy axis of 38.21$\pm$0.01$\mathrm{^o}$, showing the synchronised reversal. The field was increased (by 25~$\muup$T) to just above the switching field. The stray field magnitude is presented on a logarithmic scale. The last panel depicts the equilibrium configuration.
        }
      \label{fig:1d_reversals}
\end{figure*}

The astroid curves in Fig.~\ref{fig:0d_angles}(a) cross at 45$^\mathrm{o}$ as expected for completely uncoupled islands, in agreement with the point-dipole model, where the first NN coupling is \emph{zero}.
Next, in Fig.~\ref{fig:1d}, we compare the single island results to those from the T-shaped, 1-D array of Fig.~\ref{fig:asi_geom}(b), to examine the influence of the stray field coupling from these extended islands.

The main features of the $H_s(\theta)$ astroids of the T-shaped 1-D array [Fig.~\ref{fig:1d}(a)] are similar to those of the single islands, but important details are different.
Critically, there exists a strong coupling regime at angles around 45$^\mathrm{o}$, where the $H_s(\theta)$ profiles of the two islands snaps together.
The inset to Fig.~\ref{fig:1d}(a) shows a magnified view, centred at 45$^\mathrm{o}$, along with the two $H_s(\theta)$ curves from single \emph{uncoupled} islands from Fig.~\ref{fig:0d_angles}(a).
The `noise' at angles just below the angle of peak $H_s$ is indicative of the vertical island coupling becoming weaker to the point where the exact reversal field is subject to small variations in the simulation. 
Simultaneous reversal of the islands is clearly energetically unfavourable at angles below around 36$^\mathrm{o}$ (see Supplemental Material Fig.~S4~\cite{sup}).

Coupling between islands reduces $H_s$ compared with isolated islands.
However, the effect is not symmetric about the 45$^\mathrm{o}$ geometric axis of the array, resulting in a sloped common $H_s(\theta)$ profile for both islands that peaks at some value, decreases and then diverges.
The system is maximally stable at applied field angles which allow greater application of field before reversal occurs i.e. the angle of maximum $H_s$ within the strong coupling region.
This angle occurs at 38.25$^\mathrm{o}$ for the T-shaped array, with the spacing of maximum $H_s$ alternating between 103.5$^\mathrm{o}$ and 76.5$^\mathrm{o}$, as shown by the horizontal arrows in Fig.~\ref{fig:1d}(a).
The relative angles are shown in the sketch inset to Fig.~\ref{fig:1d}(a).

The angle of maximum $H_s$ corresponds to a magnetic easy-axis of the array, one seen during \emph{field-driven} reversal and is directly related to the existence of end-states.
Observations of the reversal direction of the array confirms this interpretation of maximum $H_s$ inside and outside of the strongly coupled regime outlined above.
When the common $H_s(\theta)$ profile splits, one island reverses first and the net moment of the system rotates either clockwise or anticlockwise, depending on the sign of the field angle offset from the anisotropy axis.
When in the strongly coupled regime, the islands reverse together, within the resolution of the 25~$\muup$T field step applied.
However, when observing the time resolved response of the system in this regime, it can be seen that reversal of the two islands are synchronised \emph{only} at the angle of maximum $H_s$, creating the minimum rotation of the net magnetisation, and diverge in opposite senses when increasing or decreasing the applied field angle from this point (see Supplemental Material Fig.~S5~\cite{sup}). 

Fig.~\ref{fig:1d_reversals} shows an example of the time evolved magnetisation and stray field during reversal at the anisotropy axis of the array (the full reversal can be seen in Supplemental Material Video~S2~\cite{sup}).
This higher angular resolution simulation more accurately locates the anisotropy axis at 38.21$^\mathrm{o}$.
As for the single island, reversal is mediated by the end-states, but now each island in the dimer reverses asymmetrically due to the inter-island dipolar coupling.

The end-state phase diagram for the T-shaped array is shown in Figs.~\ref{fig:1d}(b) and \ref{fig:1d}(c).
For all end-state types, when one island magnetisation flips, the strength of the end-states in the other island is affected.
This may be seen for s-states in Fig.~\ref{fig:1d}(c) at angles around 90$^\mathrm{o}$ by a strengthening of the s-state when the vertical island reverses (indicated by a dashed line).
This effect, however, is more easily seen elsewhere.
At small negative fields, c-states (shown in blue) are induced in the vertical island by the stray field from the horizontal one; in effect, the horizontal island biases the vertical island.
Similarly, at positive fields, large regions of c-state are created in the phase diagram for the horizontal island as a result of the stray field from the vertical island, explaining the irregular $H_s(\theta)$ profile for the horizontal island.

The vertical island has a greater influence over the net anisotropy in the dimer than does the horizontal island due to the horizontal island lying in a region where the dipolar fields from the vertical island are stronger than is the opposite case.
Further evidence of the asymmetric biasing effect can be seen in the reversal of the $H_s(\theta)$ dependence of the horizontal island in the strongly coupled regime.

Magnetostatic induced bias effects have been reported in kagome arrangements,~\cite{PANAGIOTOPOULOS_PB_2016_kagome_bias} where the angle between islands is 120$^\mathrm{o}$.
However, in pinwheel arrangements, the angle between islands is 90$^\mathrm{o}$ and, consequently, the role of end-states is far greater.
The existence of the strongly coupled regime and the modification of the anisotropy axes are direct consequences of the extended nature of the islands and the end-states they support.
This highlights the importance of considering their non-point-dipole nature.
We investigate arrays formed by islands more closely approximating point dipoles in Section~\ref{sec:island_size}.
In the next sections, we examine the four-island pinwheel vertex and consider the effects of array size in the pinwheel geometry.

\section{Pinwheel vertex}
\label{sec:pinwheel}
The smallest 2-D pinwheel array of lattice points is the four-island structure shown in Fig.~\ref{fig:asi_geom}(c).
A second, equivalent configuration of opposite (2-D) chirality is obtained from the mirror image of this array.
Both arrays have an `open' centre (i.e. the extension of their long-axes do not meet at the common point in the centre of the four islands) and $C_4$ rotational symmetry, so we consider only the one pinwheel configuration shown in the figure.
The equivalently sized square lattices are shown in Figs.~\ref{fig:asi_geom}(f) and \ref{fig:asi_geom}(g) and have a `closed' and `open' centre, respectively.

\begin{figure}[hbt]
  \centering
      \includegraphics[width=8.5cm]{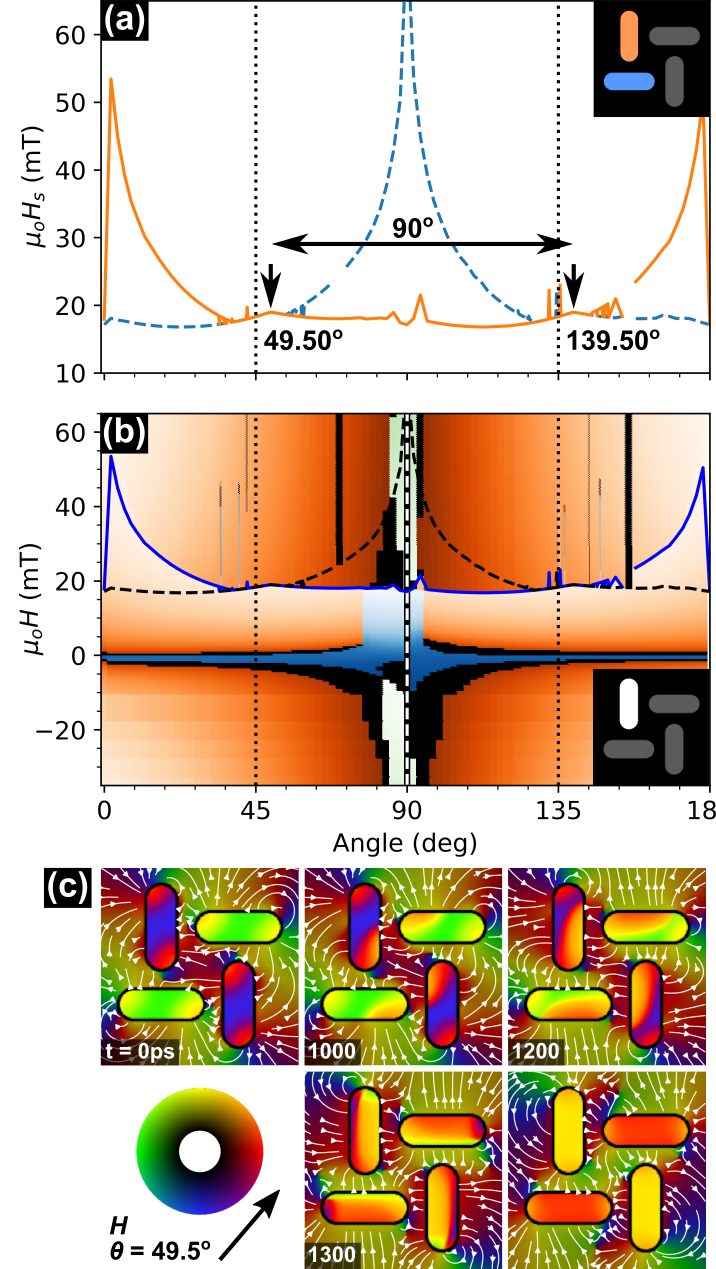}
      \caption{
        (a) $H_s$ astroids and (b) end-state phase diagram for selected islands in a pinwheel vertex. The colour maps in the phase diagram are the same as those presented in Fig.~\ref{fig:0d_angles}(c). The solid line in the phase diagram is the superimposed $H_s(\theta)$ profile of the highlighted island. The dashed line is for the other island highlighted in (a). (c) Pinwheel magnetisation and stray field during reversal at the anisotropy axis after increasing the field (by 25~$\muup$T) to just above the switching field. The stray field magnitude is presented on a logarithmic scale. 
      }
      \label{fig:pinwheel}
\end{figure}

Fig.~\ref{fig:pinwheel} summarises the results for the pinwheel array.
Both the pinwheel and square structures have $C_4$ rotational symmetry, but only the pinwheel vertex has no in-plane mirror symmetry, resulting in it having a simpler set of magnetic configurations.
The $H_s(\theta)$ astroids of both the horizontal island pairs and the vertical islands pairs in the pinwheel geometry [Fig.~\ref{fig:pinwheel}(a)] are identical and the pairs are offset from one another by 90$^\mathrm{o}$, as expected from the symmetry of the geometry.
The same is true of the end-state phase diagrams, so we show only one in Fig.~\ref{fig:pinwheel}(b).

As in the T-shaped array, a misalignment of the anisotropy axes is present in the pinwheel vertex, but the offset from 45$^\mathrm{o}$ is somewhat smaller and the anisotropy axis is offset in the opposite direction, to $>$45$^\mathrm{o}$ (49.50$^\mathrm{o}$).
Due to the rather weak coupling, leaf-states are still present in the pinwheel vertex, and the phase diagram resembles that of the horizontal island in the T-shaped geometry after applying a 90$^\mathrm{o}$ shift, but with subtle differences, such as the centre of the c-state region being offset to angles $<$90$^\mathrm{o}$.

The time resolved reversal at the anisotropy axis for the pinwheel vertex is shown in Fig.~\ref{fig:pinwheel}(c) (the full reversal can be seen in Supplemental Material Video~S3~\cite{sup}).
Reversal of all islands happens simultaneously, with the end-states closest to the centre of the vertex growing while the outer end-states shrink slightly in extent.

The results from the square arrays are shown in Supplemental Material Fig.~S6 along with a more detailed discussion of their properties.~\cite{sup}
Here, we simply note that, because the islands meet at an extension of the their long-axis, the stray field coupling in square arrays is stronger and there are no leaf-states and many more c-states as a result.
Importantly, however, because the square arrays have mirror symmetry, the anisotropy axes are aligned with the geometrical ones at 45$^\mathrm{o}$.

\section{Anisotropy mechanism}

\begin{figure*}[ht]
  \centering
      \includegraphics[width=18.0cm]{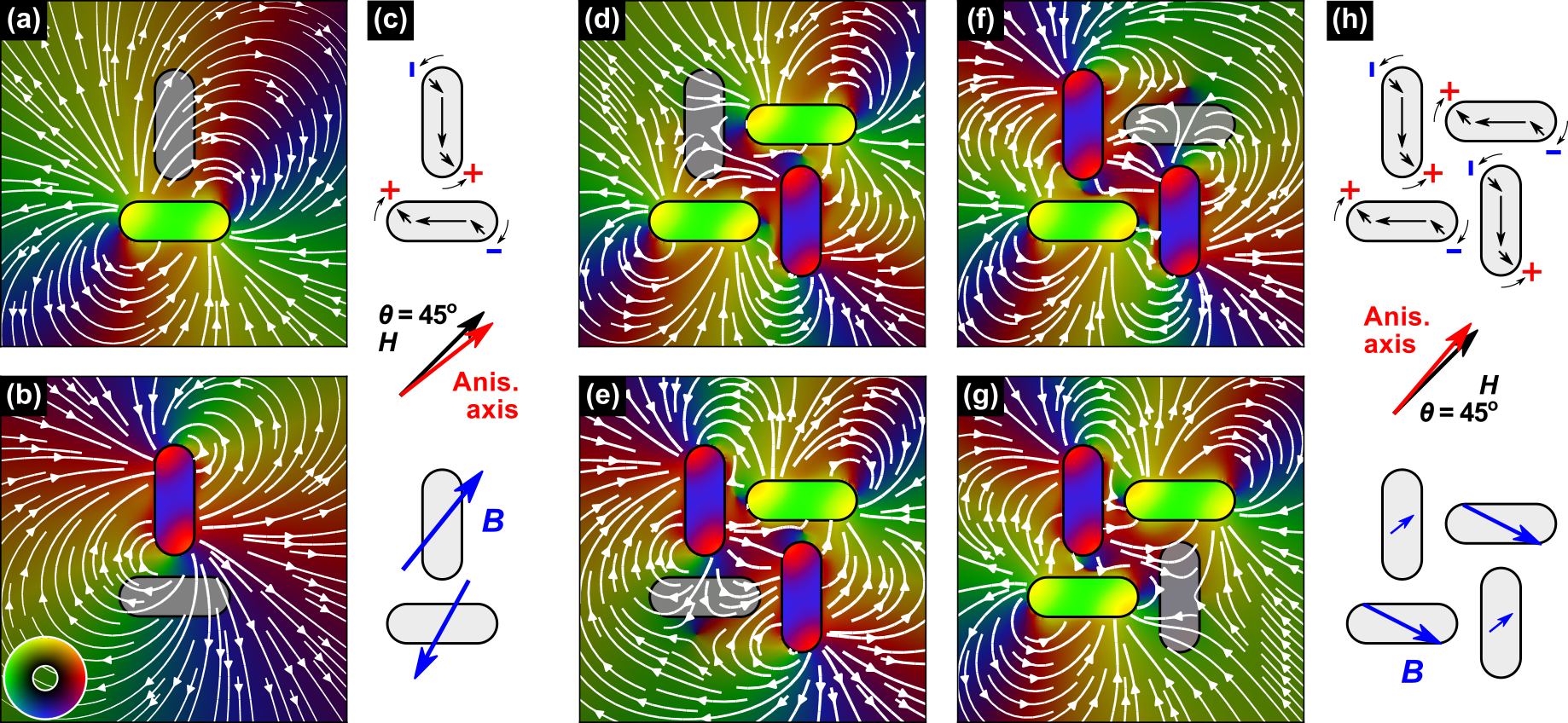}
      \caption{
            Magnetisation and stray field maps just before reversal in all configurations of (a) and (b) the T-shaped array and (d)-(g) the pinwheel vertex with one island (grey) removed. The end-state configurations are sketched in the top panels of (c) and (h), while the average induction from all islands (excluding the removed one) in the area of the removed island is shown by the blue arrows in the bottom of the same panels. The applied field angle is 45$^\mathrm{o}$. The red arrows indicate the direction of the anisotropy axes, with the angles drawn to scale. The stray field is plotted on a logarithmic scale and uses the same colour wheel as the magnetisation, shown in the inset to (b). 
      }
      \label{fig:stray45}
\end{figure*}

To understand the origin of the anisotropy axis misalignment in both the T-shaped and pinwheel arrays, we consider the magnetisation configuration just before reversal with an external field applied along the geometrical axis, at 45$^\mathrm{o}$, so that equal components lie along the long-axis of each island.
We examine the simpler case of the T-shaped array first.

Canting of the moment of each island in the dimer occurs in order to minimise the magnetisation component antiparallel with $H$, as shown in the top panel of Fig.~\ref{fig:stray45}(c).
The field from each island just before reversal is shown separately in Figs.~\ref{fig:stray45}(a) and \ref{fig:stray45}(b).
The space left by the removal of an island is marked by a grey area.
The induction ($B$) field lines in the area of the removed island are shown by the stream lines, which show the flux `flow'.
The induced field clearly lies in different directions for the two islands.
The average direction and relative strength of $B$ in the area of each missing island is marked by the blue arrows in the lower panel of Fig.~\ref{fig:stray45}(c).
For the top island, the stray field from the bottom one adds to applied field, while the opposite happens for the bottom island, with the net result that the external field must be applied at an angle $<$45$^\mathrm{o}$ for the islands to reverse together.
This mirrors the situation found in continuous film ferromagnets where anisotropies create `effective fields' which contribute the energy landscape.
In pinwheel ice, the effective field is an induction field created from the magnetisation of anisotropic islands in a constrained geometry.

The equivalent stray field and end-state plots for the pinwheel vertex are shown in Figs.~\ref{fig:stray45}(d) - \ref{fig:stray45}(h).
The situation is similar to that of the dimer, except that the canting of the island dipoles [top panel of Fig.~\ref{fig:stray45}(h)] creates virtual antivortices near to the missing islands [Figs.~\ref{fig:stray45}(d)-\ref{fig:stray45}(g)].
The stray field is plotted on a logarithmic scale and is strongest at the ends of the islands nearest to the centre of the array and it is here that the islands first begin to reverse, as shown in Fig.~\ref{fig:pinwheel}(c).
Multiple metrics for the $B$-field strength may be chosen and, in the bottom panel of Fig.~\ref{fig:stray45}(h), we represent the average value over the missing island by the blue arrows.
This metric is somewhat similar to the field within the island at the point of first reversal and explains why the anisotropy axis lies at an angle $>$45$^\mathrm{o}$ for the pinwheel vertex.
Specifically, the $B$-field from all other islands along the easy-axis of the horizontal islands is stronger than that for the vertical islands and, thus, $H$ must be applied at $\theta > 45^\mathrm{o}$ for the arrays to reverse synchronously.

Several important points arise from this analysis.
The first is that the end-states determine the field-driven anisotropy axes.
Consequently, the shape of the islands (as well as other array and island parameters) will influence the degree of inter-island coupling and, hence, the emergent array properties.
Secondly, the anisotropy axes are different for the different array shapes.
The T-shaped array has mirror symmetry (in both 2-D and 3-D) and the pinwheel vertex has an inversion centre (in both 2-D and 3-D), so the geometry is \emph{achiral}.
The magnetisation, however, breaks the symmetry, removing any rotation axis and inversion centre.
A mirror plane exists in the simulations, so the magnetisation may be regarded as \emph{2-D} chiral.
However, this is true of all of the geometries considered, including the single island and square geometry and, therefore, the misalignment of the anisotropy axes from the geometrical ones are a result of symmetry reduction---correlated interactions specific to the pinwheel geometry---but are not intrinsically a chiral effect (see Supplemental Material Fig.~S8 for further discussion of the structures from a symmetry perspective~\cite{sup}).

Another important point is that the angle between the anisotropy axes in the pinwheel vertex is 90$^\mathrm{o}$ [Fig.~\ref{fig:pinwheel}(a)].
This is a consequence of the array having a symmetric edge (c.f. Figs.~\ref{fig:asi_geom}(d) and \ref{fig:asi_geom}(e)), and is markedly different than that for the T-shaped asymmetric array shown in Fig.~\ref{fig:1d}(a), where the angle between the anisotropy axis alternates between $<$90$^\mathrm{o}$ and $>$90$^\mathrm{o}$, corresponding to a combination of cubic and uniaxial (and higher order) anisotropy contributions.
The cubic contribution creates the offset from the geometrical axes but retains the $C_4$ symmetry, while the uniaxial contribution splits the 90$^\mathrm{o}$ spacings between anisotropy axes, reducing the astroid symmetry to $C_2$.
Consequently, depending upon along which geometrical axis a pinwheel ASI array is being driven, arrays of different sizes and shapes or even the \emph{same} size and shape can appear to have different anisotropy axes.
We examine the array size dependence of this effect in the next section.

\section{Array size dependence}
\label{sec:larger2d}
\begin{figure}[ht]
  \centering
      \includegraphics[width=8.5cm]{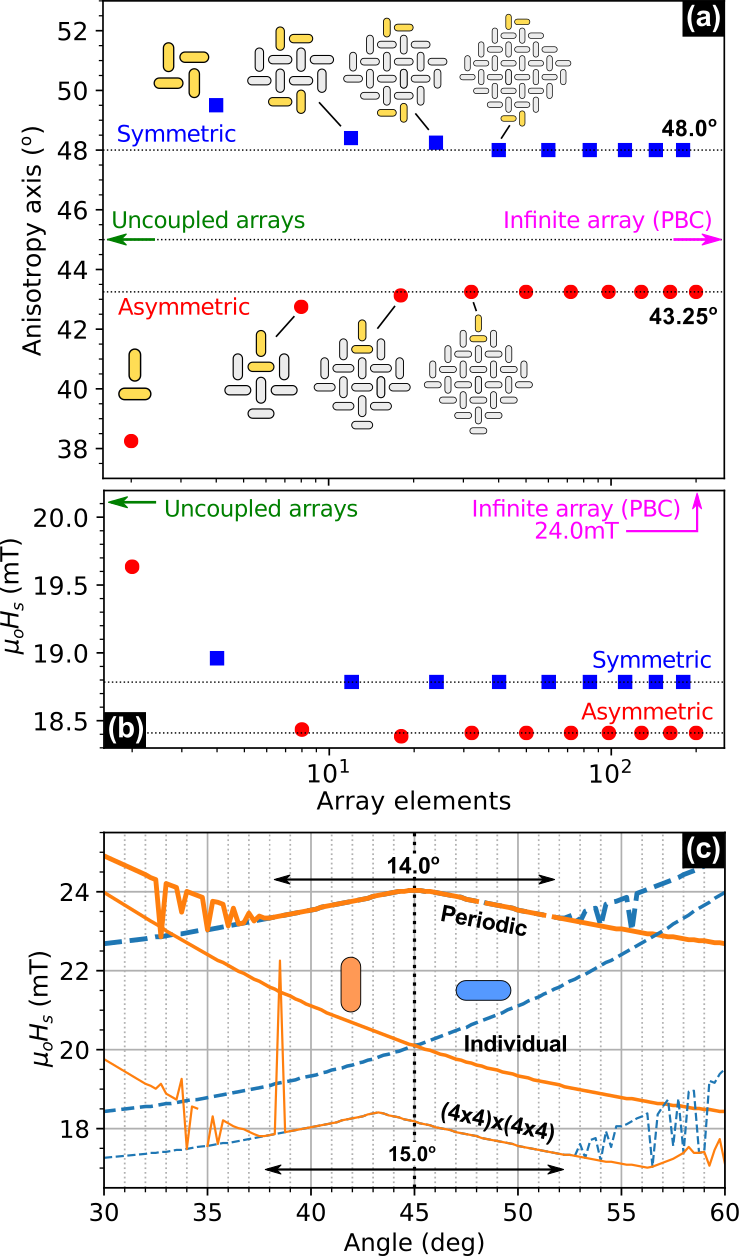}
      \caption{
        (a) The angle of the anisotropy axis of pinwheel geometries with symmetric and asymmetric edges as a function of the number of islands. For clarity, only the first few island shapes are drawn. (b) Dependence of switching field on array size. The resolution of the simulations are 0.25$^\mathrm{o}$ and 25~$\muup$T. Periodic boundary conditions (PBC) were employed for the infinite array. Five repeats on each side in $x$ and $y$ of a 64 island pattern was used, with the split down the centres of the islands at the array edges. (c) $H_s(\theta)$ astroids of the 4$\times$4 array, uncoupled islands, and the PBC array in a small angular range centred on $\theta = 45^\mathrm{o}$. The yellow islands in (a) are the the first to reverse. The arrows in (a) and (b) are for (green) an uncoupled array (identical to single islands) and (magenta) an infinite array.
      }
      \label{fig:size_anisotropy}
\end{figure}

The size and edge shape of pinwheel ASI arrays are coupled and both may influence the misalignment of the magnetic anisotropy axes (c.f. Figs.~\ref{fig:1d}(b) and \ref{fig:pinwheel}(a)).
In order to understand how the anisotropies present change with array size, we map the angle of the maximum $H_s$ value in the strongly coupled regime at angles around 45$^\mathrm{o}$ as a function of array size.
The results of this are shown in Fig.~\ref{fig:size_anisotropy}(a) for arrays ranging in number of islands from 2 to 200.
Rather than the misalignment magnitude continuously decreasing to 0$^\mathrm{o}$ as the arrays become larger, as might be predicted if the arrays were becoming bulk-like in the relevant feature, it decreases to different non-zero plateau values, depending on the array symmetry.
The anisotropy plateau angle reaches 43.25$^\mathrm{o}$ for asymmetric arrays and 48.00$^\mathrm{o}$ for symmetric arrays.

A similar effect is seen in the magnitude of $H_s$ at the anisotropy axis, as shown in Fig.~\ref{fig:size_anisotropy}(b).
The $H_s$ value is a measure of the local energy barrier to the reversal of the single island that nucleates the array reversal.
For both array-edge types, $H_s$ tends towards a plateau value, with arrays with lower symmetry, i.e. those with an asymmetric edge, tending towards a lower $H_s$ plateau faster than do the symmetric-edged arrays.

\begin{figure*}[ht]
  \centering
      \includegraphics[width=18.0cm]{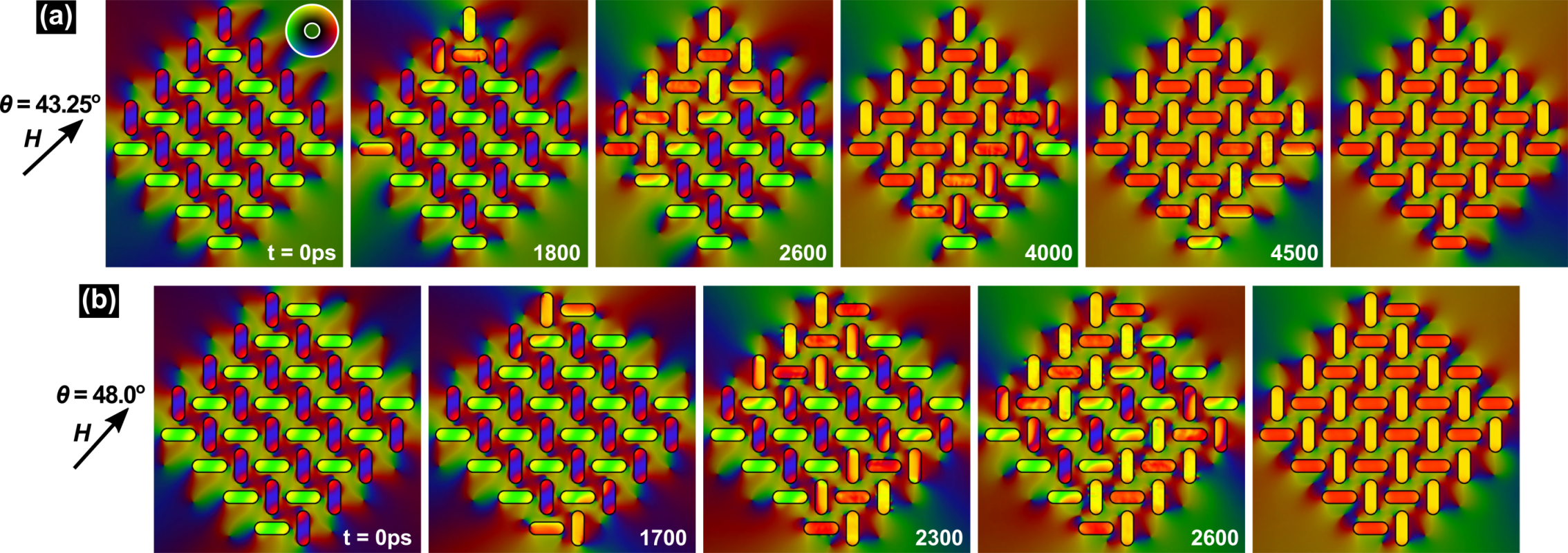}
      \caption{
            Time evolution of the magnetisation and stray field during reversal of the smallest (a) asymmetric (4$\times$4)$\times$(4$\times$4) and (b) symmetric (5$\times$4)$\times$(4$\times$5) arrays to show asymptotic characteristics at their respective anisotropy axes. The field was increased (by 25~$\muup$T) to just above the switching field. The stray field magnitude is presented on a logarithmic scale. The last panels in each row depicts the equilibrium configuration after reversal. The arrows show the applied field direction.
      }
      \label{fig:array_reversal}
\end{figure*}

The above observations can only be explained by the array reversal mechanism being mediated by islands at the array edges and, in particular, their corners where the symmetry is lowest and the barrier to reversal of an island is lower.
The first islands to reverse are marked in yellow in Fig.~\ref{fig:size_anisotropy}(a), and are indeed located at the corners of the arrays.

To confirm the importance of the array edges, we performed simulations with periodic boundary conditions (PBC).
Fig.~\ref{fig:size_anisotropy}(c) compares the $H_s(\theta)$ astroids for a vertical and horizontal island in a 64 island array with PBC [thick lines, top] with those from a finite array [thin lines, bottom] and from two uncoupled islands [medium lines, middle].
While an infinite array is somewhat impractical to realise
\footnote{%
The hairy ball theorem precludes mapping the 2-D arrays onto the surface of a sphere with the ability to apply a uniform `in-plane' field. However, uniform fields could easily be applied to arrays mapped onto a cylinder where array corners are absent and there are only two parallel edges. A torus could be used to remove edges and corners, but a very large radius would need to be used to minimise inter-island spacing non-uniformities and it is not clear how a uniform field could be generated. Measurement of the magnetic configuration of these 3-D structures would not be trivial.}, it does serve to highlight the importance of the array corners.
A strongly coupled regime also exists in the PBC simulations, over a similar angular range as that in the finite arrays (14$^\mathrm{o}$ \emph{versus} 15$^\mathrm{o}$, as shown in Fig.~\ref{fig:size_anisotropy}(c)), but the anisotropy axis lies at \emph{exactly} 45$^\mathrm{o}$.
The angle of maximum $H_s$ in the PBC array is also indicated by the magenta arrow in Fig.~\ref{fig:size_anisotropy}(a).
In the infinite array, the symmetry is restored and there are no array corners to mediate reversal, leaving the anisotropy axis aligned with the geometrical one.
The switching field for the PBC simulation is shown by the magenta arrow in Fig.~\ref{fig:size_anisotropy}(b) and is much larger than that of uncoupled arrays (green arrow) due to self-stabilisation.

The different anisotropy axes and $H_s$ plateau values in the finite arrays are due to the different edge symmetries, specifically, different array edge first and higher NN configurations.
This conclusion is consistent with previous suggestions of there existing an array-edge-dependent modification of the anisotropy axes.~\cite{Li_acs_nano_2019}
In that work, however, different anisotropy axis angles were measured for the two edge types than are found in the simulations in this work.
This difference may be due to a number of reasons, including the influence of the island shapes, which have been shown to affect the reversal process in square ice,~\cite{Kohli_PRB_2011_sq_edge_anis} and possible curvature of the membranes on which the islands were formed.
Supplemental Material Fig.~S7 and the related discussion expands on the role of island shape in determining the anisotropy axis in more detail.~\cite{sup}

The plateau in anisotropy axes is first reached in the pinwheel array of 32 islands, formed by two interleaved 4$\times$4 subarrays, which is thus the smallest array in which the mediating island has an effectively `full' set of NNs.
Consequently, the maximum relevant scale of the inter-island interactions must be $\le$~7 lattice units.
We investigate the reversal process and inter-island coupling during it in more detail in the next section.

\section{Array reversal process}
\label{sec:reversal_mech}
Fig.~\ref{fig:array_reversal} shows the array reversal pattern for the 4$\times$4 asymmetric array (top row, panel (a)) and the 5$\times$4 symmetric array (bottom row, panel (b)) with the applied field increased to just above $H_s$ (the full reversal is shown in Supplemental Material Videos~S5 and S6~\cite{sup}).
For the asymmetric array [Fig.~\ref{fig:array_reversal}(a)], nucleation begins at the island with the lowest configuration of NN, the vertical island of the T-shaped subarray in the top corner, and propagates through an avalanche of first NN reversals.
This results in a 2-D reversal through mesoscopic domain wall propagation perpendicular to the applied field direction, with the bottom island the last one to reverse. 
The last island to reverse is also in a low symmetry environment, but oriented perpendicular to the first island to reverse.

The main features of the simulation compare well to the reversal process seen in the experiment previously reported.~\cite{Li_acs_nano_2019}
The experimental data~\cite{Li_acs_nano_2019_enlighten} was obtained from Fresnel imaging Ni$_{80}$Fe$_{20}$ arrays in a transmission electron microscope and is compared against the simulations in detail in Supplemental Material Fig.~S10.~\cite{sup}
Fig.~\ref{fig:4444_rev_coherence}(a) gives one example of the experimental array reversal pattern for the asymmetric array at the experimentally determined anisotropy axis, mapped using the island switching field.
In the experiment, array reversal nucleates at the corners or edges of the array, and FM mesoscopic domain growth is through mesoscopic domain wall formation and propagation perpendicular to the applied field direction.
In both simulation and experiment, the reversal direction was always the same as that in Fig.~\ref{fig:array_reversal}(a), from top right to bottom left, irrespective of the field sweep direction (c.f. Supplemental Material Figs.~S10(a) and S10(b)~\cite{sup}), due to the reduced array symmetry discussed earlier.
The main difference in the experiment is that imperfections in the real system cause the reversal to be through several cascades spread out over multiple field steps, as indicated by the facets of uniform colour in Fig.~\ref{fig:4444_rev_coherence}(a). 

Similar nucleation behaviour to that discussed here has been observed in square ASI, where Dirac strings nucleate from the edges and corners of the arrays.~\cite{Phatak_PRB_2011_sq_monopole_edge}
The array edges have also been observed to be important in determining the array dynamics in ideal square systems due to limitations on the vertex type nucleation location,~\cite{Budrikis_PRL_2010_sq_dynamics} but imperfections in real systems wash out its effect.~\cite{budrikis_PRL_2012_sq_disorder}
In the pinwheel geometry, the near degenerate vertex energies,~\cite{Macedo_PRB_2018} the existence of the strong coupling regime, and the large difference in the switching field in the centre of arrays compared to at the edges and corners (c.f. PBC with finite arrays in Fig.~\ref{fig:size_anisotropy}(c)) reduce the influence of disorder in the system compared with square ice.

The reversal pattern for the symmetric array [Fig.~\ref{fig:array_reversal}(b)] is similar but, here, the T-shaped dimers where reversal nucleates (marked in yellow in Fig.~\ref{fig:size_anisotropy}) are rotated by 90$^\mathrm{o}$ with respect to the asymmetric array and, due to the array symmetry, there are two nucleation points, resulting in the two mesoscopic domain walls propagating towards the centre of the array.
The measured reversal pattern from Fresnel transmission electron microscopy of a symmetric array is shown in Supplemental Material Fig.~S12~\cite{sup} and shares the same general trend as that seen in the simulations, with some variability in the exact reversal pathway.

\begin{figure*}[ht]
  \centering
      \includegraphics[width=17.0cm]{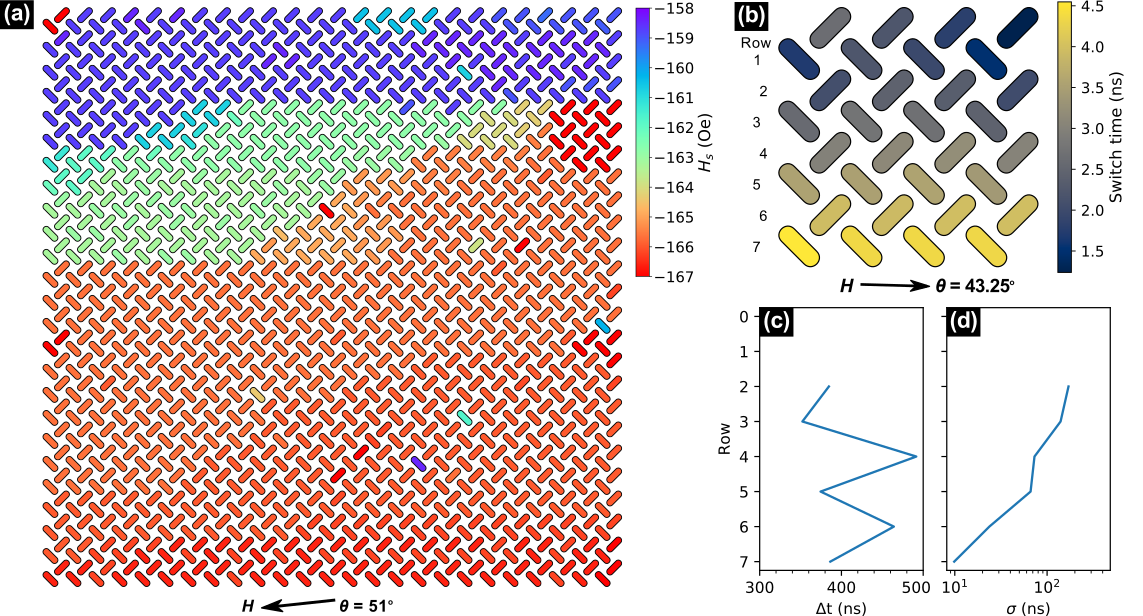}
      \caption{
            (a) Experimental reversal pattern for a large asymmetric array at the anisotropy axis during a decreasing field sweep. (b) Island switching time map for simulations of the asymmetric (4$\times$4)$\times$(4$\times$4) array in Fig.~\ref{fig:array_reversal}(b), during an increasing field sweep. (c) Average time between row reversals and (d) standard deviation in switching time along rows of the array in (b). The data in (c) and (d) are from islands in the third column to the last column of (b). The array is rotated clockwise by 45$^\mathrm{o}$ for display. 
      }
      \label{fig:4444_rev_coherence}
\end{figure*}

To map the reversal in the micromagnetic simulations in more detail, we plot in Fig.~\ref{fig:4444_rev_coherence}(b) the time of switching of each island in the asymmetric array of Fig.~\ref{fig:array_reversal}.
It is clear that the mesoscopic DW propagation direction is perpendicular to the applied field.
However, small changes in the applied field angle can significantly change the angle of the reversal.
For example, with the 45$^\mathrm{o}$ field, the reversal avalanche propagates along the diagonal direction of the array (as shown in Supplemental Material Fig.~S11~\cite{sup}).

Fig.~\ref{fig:4444_rev_coherence}(c) shows a measure of the time between reversal of adjacent rows as the 2-D domain extends downwards.
We omit the first few rows in the plot, where there is some component of mesoscopic domain growth horizontally, antiparallel to the applied field.
The average velocity of the mesoscopic domain wall is 514~ms$^{-1}$.
This is very similar to the velocity of the magnetisation component of each island along the mesoscopic domain wall propagation direction of 512$\pm$25~ms$^{-1}$.
The comparable speeds, without significant reduction in the mesoscopic wall speed, is evidence of strong coupling between first NN islands.
Fig.~\ref{fig:4444_rev_coherence}(d) plots the standard deviation of the switching time across the rows as the domain grows, and shows that synchronisation of the DW increases exponentially as a result of the collective dipolar inter-island interactions.

The $H_s$ values of all islands in simulations of large but finite arrays of a given symmetry are the same as the corner islands because, once the first island reverses, inter-island coupling results in an avalanche of island reversal across the array.
This and the other features described above are strongly suggestive of first NN coupling being dominant in our field-driven pinwheel array simulations and experiments.
For example, the angular range of the strongly coupled regime remaining constant irrespective of array size [Fig.~\ref{fig:size_anisotropy}(c)] is indicative of a localised interaction; it is not related to the array edge or symmetry.
Indeed, previous experiments on a much larger array~\cite{Li_acs_nano_2019} showed that 2-D FM behaviour was limited to a similarly small range of angles.

To confirm if first NN coupling is dominant, we examined the induction field in regions located at the point of reversal for an island of each orientation towards the centre of the (4$\times$4)$\times$(4$\times$4) array and did indeed see that the field only increases significantly when the first NN islands reverse (see Supplemental Material Fig.~S13 and related discussion~\cite{sup}).

The emergence of first NN coupling with the application of field and the associated superferromagnetism can in some ways be regarded as analogous to a Heisenberg exchange interaction in continuous ferromagnets, but one that \emph{only} exists under an applied field.
In this context, the array edge dependence of the array reversal may be regarded as a result of a different surface spin-states of the superferromagnets.
Indeed, it is interesting to observe the similarities in the reversal patterns of the superferromagnet with a symmetric edge [Fig.~\ref{fig:array_reversal}(b)] to that of the single island [Fig.~\ref{fig:0d_mh}(b)].
In both structures, reversal begins at the `corners' and symmetrically extends along one axis of the structure, with the centre the last section to reverse.

First NN coupling is dominant in pinwheel reversals when in the strongly coupled regime and, in fact, is responsible for the \emph{existence} of the strongly coupled regime, as demonstrated by the coupling in the T-shaped array.
Outside the strongly coupled regime, first NN coupling is also very significant (see Supplemental Material Fig.~S14~\cite{sup}) but the domains that form during field-driven reversals are more irregular.

\begin{figure}[ht]
  \centering
      \includegraphics[width=8.5cm]{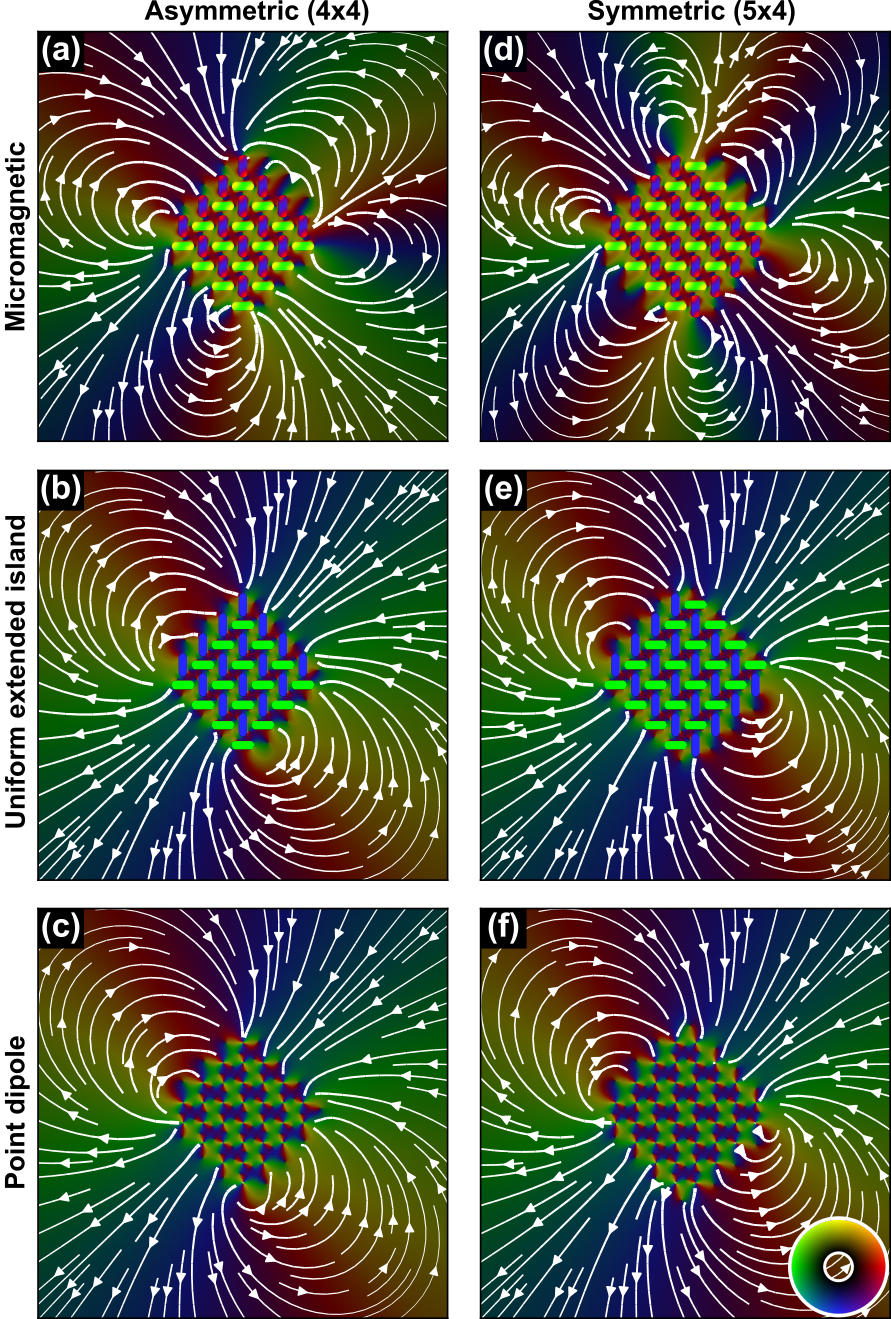}
      \caption{
            Magnetisation and stray field for the (4$\times$4)$\times$(4$\times$4) (left column) and (5$\times$4)$\times$(4$\times$5) (right column) arrays just before reversal for different models (in rows): (a) and (d) micromagnetic, (b) and (e) uniformly magnetised extended island, and (c) and (f) point dipole; showing the strong influence of the end-states on the stray field distribution. The magnetisation and stray field configurations in (a) and (d) are identical to those in the first columns of Figs.~\ref{fig:array_reversal}(a) and \ref{fig:array_reversal}(b), respectively. The stray field magnitude is presented on a logarithmic scale. The colour map is shown in the inset to (f).
      }
      \label{fig:stray_field}
\end{figure}

To further demonstrate the importance of dipolar fields from physically extended islands, we plot in Fig.~\ref{fig:stray_field} the magnetisation and stray field distribution for the asymmetric and symmetric arrays shown in Fig.~\ref{fig:array_reversal} at the point just before reversal, but over a larger area.
Three models are considered: micromagnetic is shown in the first row; extended islands each with a uniform magnetisation along their long-axes, i.e. with no end-states, in the second; and a point-dipole model in the third.
Outside the array, the stray field from the uniformly magnetised extended islands and the point-dipoles is very similar for each array type and, indeed, across array types, and resembles that of a dipole.
For the micromagnetic case, however, the end-states dramatically change the stray field distribution, creating complex lobed patterns unique to each array symmetry, and that ultimately influences the collective behaviour of the arrays \emph{and} the field outside the arrays.

Another important factor that may be used to tune the array properties is the island size, and we investigate this next.

\section{Towards Point Dipoles}
\label{sec:island_size}
In field-free environments, the ground state in pinwheel ice has been shown to have FM ordering due to long range dipolar coupling.~\cite{Macedo_PRB_2018}
In this case, the point-dipole model and micromagnetic calculations give similar results.
A related 2-D ASI system with four states per island and configurable array magnetic ordering has also shown good agreement between the point-dipole and micoromagnetic models when considering the GS configuration.~\cite{Louis_NatMat_2018_potts}
The results of the previous sections show that a very different mechanism---end-state induced pseudo-exchange from dipolar coupling between first NN islands---gives rise to FM ordering in field-driven experiments on pinwheel structures.~\cite{Li_acs_nano_2019}  
Here, we consider how the magnetic ordering of field-driven pinwheel arrays changes as the islands are reduced in their in-plane size and they increasingly approximate point-dipoles.

\begin{figure}[ht]
  \centering
      \includegraphics[width=8.5cm]{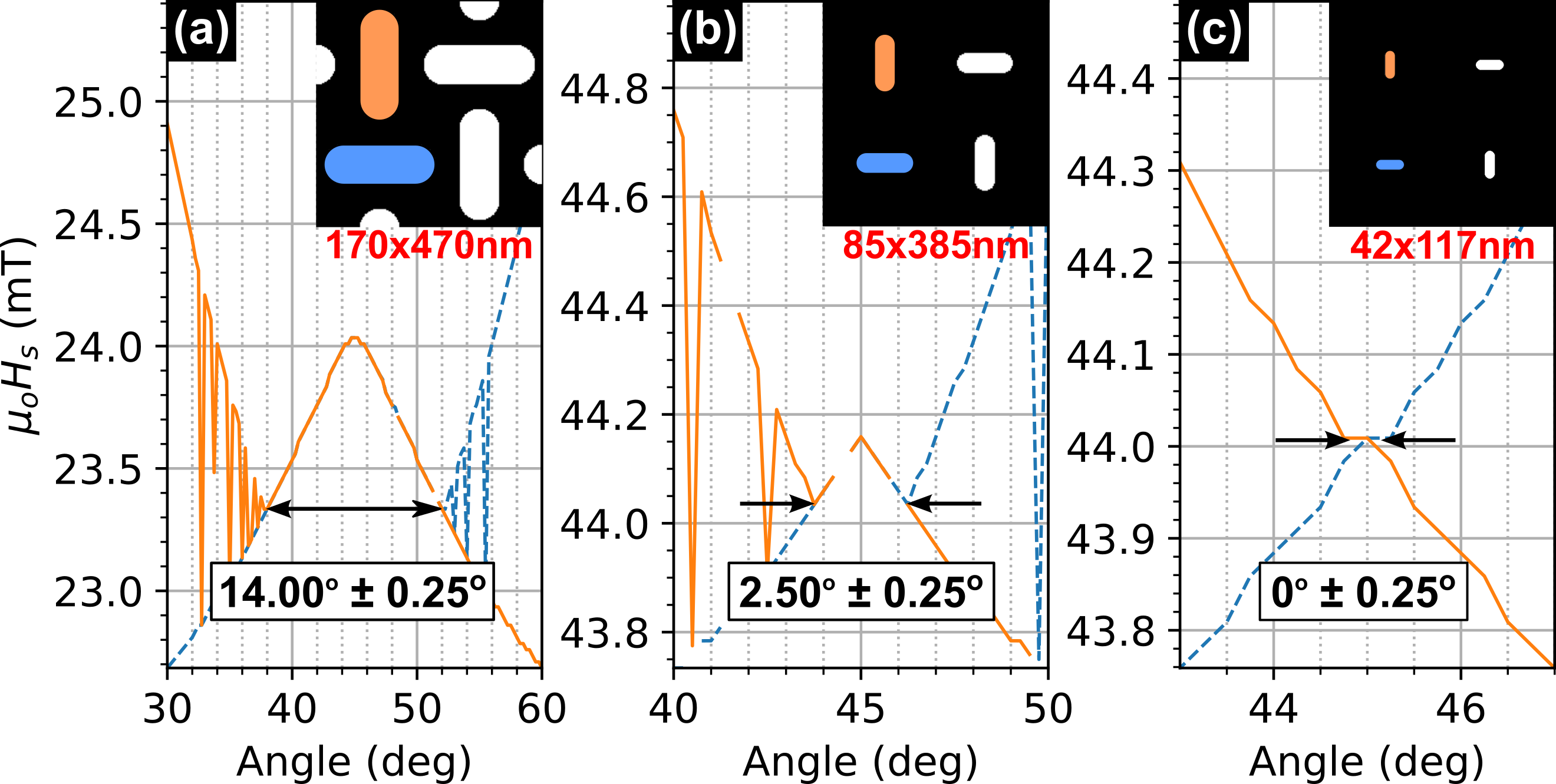}
      \caption{
            $H_s(\theta)$ astroids for periodic arrays of islands of different in-plane sizes (a) original, (b) half, (c) quarter, all on lattices of the original spacing, showing reduction in the angular range of the strongly coupled regime. The angular resolution was 0.25$^\mathrm{o}$. The insets show example islands on a common scale. The in-plane grid size of the simulations was 2.5~nm in (c) and 5~nm otherwise.
      }
      \label{fig:pbc_island_size}
\end{figure}

To investigate the effect of island size, we performed simulations with islands of reduced in-plane size while maintaining the island thickness and 2-D shape, and the lattice constant of the array.
Periodic boundary conditions were used to remove the influence of the array edges.
We note that, with this approach, multiple inter-dependent parameters are changing as the islands are reduced in size.
These include the net moment of each island, the inter-island edge-to-edge spacings, and the island edge curvature with respect to the exchange length.
Of these, the important variables affecting the coupling are the inter-island edge spacing and the exchange length compared to the island size.
Ultimately, however, the simulation sequence does show how the array properties evolve as the islands increasingly approximate point dipoles.

The results of these simulations are shown in Fig.~\ref{fig:pbc_island_size}, where it can be seen that the width of the strongly coupled regime reduces from 14.00$^\mathrm{o}$ for the nominal island size [Fig.~\ref{fig:pbc_island_size}(a)] to less than the 0.25$^\mathrm{o}$ resolution of the simulations for islands one quarter of the in-plane size [Fig.~\ref{fig:pbc_island_size}(c)], as the islands more closely resemble point dipoles.

The existence of the strongly coupled regime is associated with 2-D FM reversal and the narrowing of the angular range of the regime is clear evidence that the islands are becoming less strongly coupled.
For quarter sized islands in the PBC array, outside the very small ($<$0.25$^\mathrm{o}$) angular range where the switching fields of the two subarrays overlap, the islands are not sufficiently coupled for 2-D reversal to occur across both subarrays and, instead, one subarray will completely reverse before the other one begins.
This is in marked contrast to point-dipole models where the net island moment will affect the vertex energy spacing, but not their ordering and thus whether the ground state of pinwheel arrays is ferromagnetic.

With the same quarter island size, the strongly coupled regime of the asymmetric array formed by two interleaved 4$\times$4 subarrays reduces in width from 15.0$^\mathrm{o}$ [Fig.~\ref{fig:size_anisotropy}(c)] to around 1.5$^\mathrm{o}$ (see Supplemental Material Fig.~S15~\cite{sup}) and $H_s$ now peaks at an angle of approximately 44.75$^\mathrm{o}$; the array edge related anisotropy is effectively removed.
Relatively homogeneous 2-D reversal still occurs at applied field angles of exactly 45$^\mathrm{o}$ but with the DW propagation direction lying along the array diagonal, as shown in Fig.~\ref{fig:4444small_array_reversal}(a) (the full reversal can be seen in Supplemental Material Video~S7~\cite{sup}), demonstrating that inter-subarray coupling still has some influence at this island size.
Note that each island in Fig.~\ref{fig:4444small_array_reversal} is drawn 3$\times$ the real size for clarity; the same data with the islands plotted to scale are shown in Supplemental Material Fig.~S16.~\cite{sup}

\begin{figure}[ht]
  \centering
      \includegraphics[width=7.0cm]{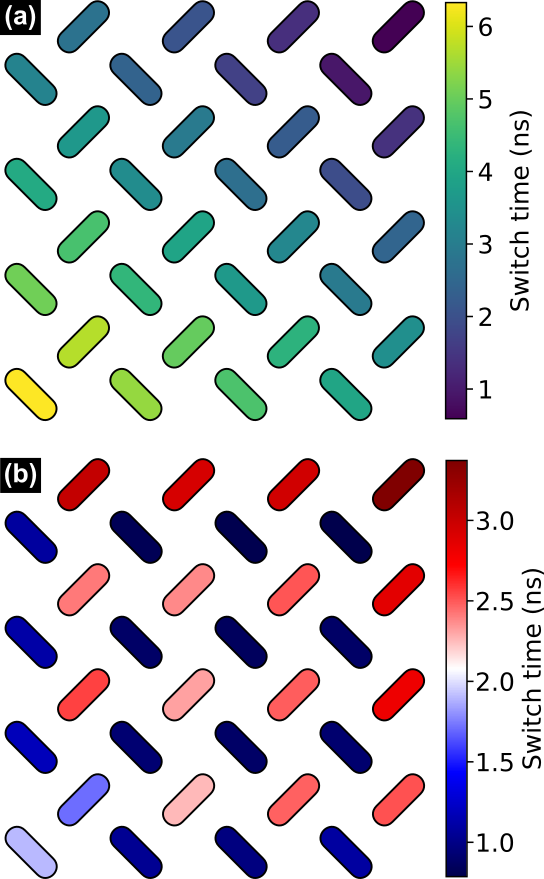}
      \caption{
            Island switching time during time evolution of the asymmetric array formed by two interleaved 4$\times$4 subarrays of quarter sized islands (42~nm $\times$ 117~nm $\times$ 10~nm), at applied field angles of (a) 45$^\mathrm{o}$ and (b) 44$^\mathrm{o}$. The field was increased (by 25~$\muup$T) to just above the switching field in (a), and was increased at a rate of 1~$\muup$T/10~ps in (b) from two different points in the field sweep. The arrays are rotated 45$^\mathrm{o}$ clockwise for display. Each island is drawn 3$\times$ the real size for clarity. Real-scale islands are shown in Supplemental Material Fig.~S16.~\cite{sup}
      }
      \label{fig:4444small_array_reversal}
\end{figure}

Due to the reduction in first NN coupling with smaller island sizes, at angles just outside the strong coupling regime, reversal occurs over multiple field values across one subarray, instead of occurring through an avalanche of first NN coupling driven homogeneous 2-D reversal.
There does still exist significant intra-subarray coupling, which manifests in the formation of a 2-D \emph{subarray} mesoscopic domain wall which propagates across the array in opposite directions for each subarray, as shown in Fig.~\ref{fig:4444small_array_reversal}(b) (the full reversal can be seen in Supplemental Material Video~S8~\cite{sup}).
This behaviour corresponds to the formation of 90$^\mathrm{o}$ N{\'e}el walls when viewing the array in terms of vertex moments.
Similar behaviour has been seen experimentally in arrays of the nominally sized islands at applied field angles outside the 2-D FM reversal regime.~\cite{Li_acs_nano_2019}   
This work shows that this feature, and many other properties of the array, including the observation of 2-D ferromagnetism are intrinsically linked to the island size and the end-states it supports, thus, demonstrating another way in which the inter-island interaction strength in pinwheel ASI arrays may be tuned.

\section{Conclusions}
We have investigated the inter-island interactions during field-driven reversals in pinwheel ice though micromagnetic modeling.
Key to understanding the reversal behaviour is the inclusion of magnetisation end-states of the physically extended islands.
End-states significantly modify the inter-island coupling, giving rise to a Heisenberg pseudo-exchange interaction when under an external field, driving first NN coupling within a range of angles around the geometrical axis of the arrays.

Related to the strong localised coupling are emergent anisotropies, consisting of different uniaxial and cubic contributions to the energy landscape, depending on the array symmetry.
The anisotropies are misaligned with the geometrical axes, and reduce in magnitude with increasing array size, plateauing at different values from the geometrical axes.

Symmetry reduction at the array edges creates an inhomogeneous island switching field distribution which results in avalanche reversals, mediated by islands at the array corners.
Reversal occurs by the formation of a 2-D superferromagnetic mesoscopic domain which grows through propagation of 180$\mathrm{^o}$ N{\'e}el walls.
Varying the island size alters the inter-island coupling, allowing the emergent properties to be tuned.
Smaller islands reduce inter-subarray coupling, with 90$\mathrm{^o}$ N{\'e}el mesoscopic domain walls becoming more prevalent as a result.

All the above characteristics are absent in or incompletely described by the point-dipole model, and only occur when the non-Ising nature of the extended islands is taken into account.
While the exact anisotropy angles are relatively sensitive to imperfections, the general feature of 2-D FM array reversal is robust and matches well the magnetisation behaviour seen experimentally.

These insights are crucial to a full understanding of the collective behaviour of pinwheel ice arrays for use in fundamental research and in potential applications such as Hall circuits based on the anisotropic magnetoresistance effect in interconnected ASI arrays,~\cite{Chern_pra_17_hall_circuits} logic,~\cite{Imre_science_2006_ASI_computation, Arava_2018_sq_logic, Caravelli_arxiv_2018_bool_gates} and neuromorphic computation,~\cite{Jensen18_asi_computation, Arava_PRA_2019_asi_relax} where inter-island interactions may be used to modify reversal paths.

Our results may also be relevant to thermalisation experiments.
Field-driven island reversal is from one s-state to another, but the inter-island interactions also control the end-states at remanence in arrays initialised by field polarisation.
While the field that these end-states induce will be weaker than those in field-driven interactions, they will still play a role in defining the energy landscape.
Further research is needed to better understand the significance of this effect in thermal experiments.

Finally, collective spin wave modes from dipolar coupling in square ice are known to occur,~\cite{Gliga_PRL_2013_sq_fmr, Jungfleisch_PRB_2016_sq_fmr} but this has yet to be observed in pinwheel ice where the inter-island coupling is reduced.
Our results show that strong coupling can exist in pinwheel ice and suggest that it may possible to observe collective microwave dynamics in this system.

Original data files are available at DOI TBA.

\section*{Acknowledgements}
This work was supported by EPSRC via grant nos. EP/L002922/1, EP/L00285X/1, and EP/M024423/1.
G.M.M. is supported by the Carnegie Trust for the Universities of Scotland.
Y. L. received support from the China Scholarship Council.
R.M. is supported by the Leverhulme Trust.
R.L.S. acknowledges the support of the Natural Sciences and Engineering Research Council of Canada (NSERC) - R.L.S. a {\'e}t{\'e} financ{\'e} par le Conseil de recherches en sciences naturalles et en g{\'e}nie du Canada (CRSNG).
G.W.P. thanks Aurelio Hierro Rodriguez for helpful discussions on the MuMax3 package.

%

\onecolumngrid

\end{document}